\documentclass[12pt, draftcls, onecolumn]{IEEEtranTCOM}

\usepackage{amssymb,bbm}
\usepackage{color,graphicx}
\usepackage{cite}
\usepackage[cmex10]{amsmath}
\usepackage{amsopn}
\usepackage{floatrow}
\usepackage{cite}

\newtheorem{theorem}{Theorem}
\newtheorem{lemma}[theorem]{Lemma}

\newtheorem{definition}{Definition}

\newcommand\qed{\hfill \rule{1.2mm}{2.8mm}}

\newcommand{\R}{\ensuremath{\mathbb{R}}}

\newcommand{\floor}[1]{\lfloor #1 \rfloor}

\newcommand{\be}{\begin{equation}}
\newcommand{\ee}{\end{equation}}

\newcommand{\bea}{\begin{eqnarray*}}
\newcommand{\eea}{\end{eqnarray*}}

\newcommand\rr{{\mathbb R}}

\newcommand\nn{{\mathbb N}}

\newcommand{\ZO}{{\mathbb N}_0}
\newcommand{\NN}{{\mathbb N}}

\newcommand{\remove}[1]{}

\newcommand{\abs}[1]{\lvert #1 \rvert}
\newcommand{\prob}[1]{{\mathbb P}\left\{#1\right\}}
\newcommand{\expect}[1]{{\mathbb E}\left[  #1\right]}

\newcommand{\lb}{\left(}
\newcommand{\rb}{\right)}
\newcommand{\lc}{\left\{}
\newcommand{\rc}{\right\}}
\newcommand{\ls}{\left[}
\newcommand{\rs}{\right]}

\newcommand{\fstar}{f^*}
\newcommand{\xstar}{x^*}
\newcommand{\ystar}{y^*}
\newcommand{\tstar}{\tau^*}

\newcommand{\VTp}{V^{\pi}_T}
\newcommand{\NTp}{N^{\pi}_T}
\newcommand{\fpT}{f^\pi_T}
\newcommand{\Rmax}{\overline{R}}
\newcommand{\Rmin}{\underline{R}}
\newcommand{\bfa}{{\bf a}}
\newcommand{\rhoint}{[ \underline{\rho}_\ell, \overline{\rho}_\ell ]}

\newcommand{\al}{a_\ell}

\newcommand{\VLP}{V^{(LP)}_T}

\newcommand{\NTHLn}{N_{T,H}^{(\ell,n)}}

\newcommand{\ATLP}{V^{({\bf a}_T)}}
\newcommand{\rell}{\rho_\ell}

\newcommand{\xm}[1]{\lambda\left( #1 \right)}
\newcommand{\nm}[1]{\eta\left( #1 \right)}

\newcommand{\suml}{\sum_{\ell = 1}^{L}}
\newcommand{\maxl}{\max_{\ell=1}^L}
\newcommand{\calC}{\ensuremath{\mathcal{C}}}

\newcommand{\rhoellscre}{\underline{\rho}_\ell}
\newcommand{\rhoellbar}{\overline{\rho}_\ell}

\newcommand{\fbar}{\overline{f}}
\newcommand{\fellbar}{\overline{f}_\ell}
\newcommand{\Dbar}{\overline{D}}
\newcommand{\Nbar}{\overline{N}}
\newcommand{\Tbar}{\overline{T}}

\newcommand{\barsigma}{\overline{\sigma}}

\newcommand{\bartau}{\overline{\tau}}
\newcommand{\Rsellxi}{R^\sigma_\ell(\xi)}
\newcommand{\fsgma}{f^\sigma}
\newcommand{\esell}{\mu^\sigma_\ell}
\newcommand{\es}{\mu^\sigma}

\newcommand{\erell}{\mu^\varrho_\ell}

\newcommand{\rhosgmaell}{\rho^\sigma_\ell}
\newcommand{\barxsell}{\overline{x}^\sigma_\ell}
\newcommand{\rhorhoell}{\rho^\varrho_\ell}
\newcommand{\barfsgma}{\overline{f}^\sigma}
\newcommand{\Aseta}{A^\sigma_{{\boldsymbol \mu}}}
\newcommand{\Uell}{U_\ell}
\newcommand{\FellP}{\overline{F}^P_\ell}
\newcommand{\FellM}{\overline{F}^M_\ell}
\newcommand{\FzeroM}{\overline{F}^M_0}
\newcommand{\FM}{\overline{F}^M}

\newcommand{\deltf}{\delta f}
\newcommand{\trho}{\tilde{\rho}}
\newcommand{\uB}{\overline{D}}

\begin{document}

\sloppy

\title{Capacity and Stable Scheduling in Heterogeneous Wireless Networks}

\author{Stephen V. Hanly, Chunshan Liu, Philip Whiting
\thanks{The authors are with the Department of Engineering, Macquarie University, Sydney, NSW, 2109, Australia. email: \{stephen.hanly,chunshan.liu,philip.whiting\}@mq.edu.au.}
}

\maketitle

\begin{abstract}
Heterogeneous wireless networks (HetNets) provide a means to increase network capacity by introducing
small cells and adopting a layered architecture. HetNets allocate resources flexibly
through time sharing and cell range expansion/contraction allowing a
wide range of possible schedulers. In this paper we define the capacity of a HetNet down link in terms of the maximum number of downloads
per second which can be achieved for a given offered traffic density. Given this definition
we show that the capacity is determined via the solution to a continuous linear program (LP).
If the solution is smaller than 1 then there is a scheduler such that the
number of mobiles in the network has ergodic properties with finite mean waiting time.
If the solution is greater than 1 then no such scheduler exists. The above results continue to hold if a more general class of schedulers is considered.
\end{abstract}


\section{Introduction}
\label{sec_intro}
HetNets have been proposed as a means to increase the
capacity of wireless networks by introducing short range pico cells
into the existing coverage area of macro cells. Significant increase in user data
throughput is possible mainly because the pico cells can operate simultaneously, providing better time or frequency re-use. This is particulary the case if traffic is
spatially concentrated into small parts of the coverage area known
as traffic ``hot-spots''.

A brief description of HetNet operation is as follows. First, time
is divided into small duration frames known as Almost Blanking Subframes
(ABS), \cite{standards}. These are used to time share between macro cell
transmissions and pico cell transmissions.  In this way interference between
the macro cell(s) and the pico cells is avoided. Additional
flexibility is provided by allowing the pico cells to adjust their
areas of coverage, that is to expand and serve more mobiles or to
contract to serve fewer mobiles, but at potentially higher rates. This
procedure is known as Cell Range Expansion (CRE). As described in standards, CRE uses received signal strength to
determine to which Base Station (BS) a mobile should connect. To make the cell
expand a bias (on the order of 0 - 16 dB) is added to the received signal strength
to determine cell size. Under our model of fixed (at the time
of arrival) location dependent rates we show that an optimal scheme assigns mobiles on
the basis of their rate ratios.


A problem central to the operation of HetNets is how to maintain
stable operation (prevent indefinite build up of backlogged traffic) against
unknown traffic demands. Results associated with this problem
were obtained  in the following papers \cite{Andrews2012,Chen2011,Hu2011,Rudolf2012,VTC2013,ICC2013}. None of these papers address the question of
dynamic stability in conjunction with the joint HetNet ABS and
CRE controls.

In fact the very flexibility in setting ABS slots over time in
conjunction with the pico coverage areas, makes it more difficult
to determine whether a particular offered traffic density can be
supported or not. Nevertheless, we show that given an
offered traffic density (fraction of mobiles arriving in the vicinity
of a location) then there is a well defined notion of capacity. This capacity
is determined as the supremum of all arrival rates which can be supported
with finite queueing delays. This depends not only on the
traffic locations of arriving mobiles but also on the data demands made by the mobiles. In this
paper we suppose all mobiles have a single file of random but bounded length
to download. Once downloaded the mobile departs the system
and does not return. Generalizations of this assumption can be
made of course but the notion of capacity will still be upheld.

The remainder of the paper is as follows. Section
\ref{sec_model} presents our model for the HetNet network, along the lines of
\cite{ICC2013} as well as some preliminary results.  The first of these identify how to
``clear'' a given set of mobiles from the network in minimum time. The second is
a continuous analogue in which total transmission time per unit time is
minimized. This latter problem takes the form of a continuous Linear Program~(LP). (Perhaps more strictly
this should be seen as a problem in the calculus of variations with integral constraints, with bounded
integrable functions, see Section \ref{sec_model}. However we retain this more informal nomenclature for this class
of problems.) With these constructions in hand, we proceed by characterising their respective solutions.
In Section \ref{sec_capres} we show that if for a given overall arrival rate $\lambda_S$
the solution to the continuous LP is smaller than 1, then a scheduling policy exists
such that the network is stable.

Section \ref{sec_converse} demonstrates a converse result, i.e.,
the number of mobiles queueing for service must steadily build up (at linear rate, almost surely) if
the solution to the continuous LP is greater than 1.  We then go on to
look at a more general case where individual picos may be on or off
during ABS time with the effect that physical rates in on cells will be higher
as the interference is reduced. Finally, Section \ref{sec_numer}
presents some illustrative numerical experiments and in Section \ref{sec_conc} there are some brief conclusions
and suggestions for further work.

\section{Model and Preliminaries}
\label{sec_model}
\subsection{Model}
\begin{figure}
\centering
\includegraphics[width=2in]{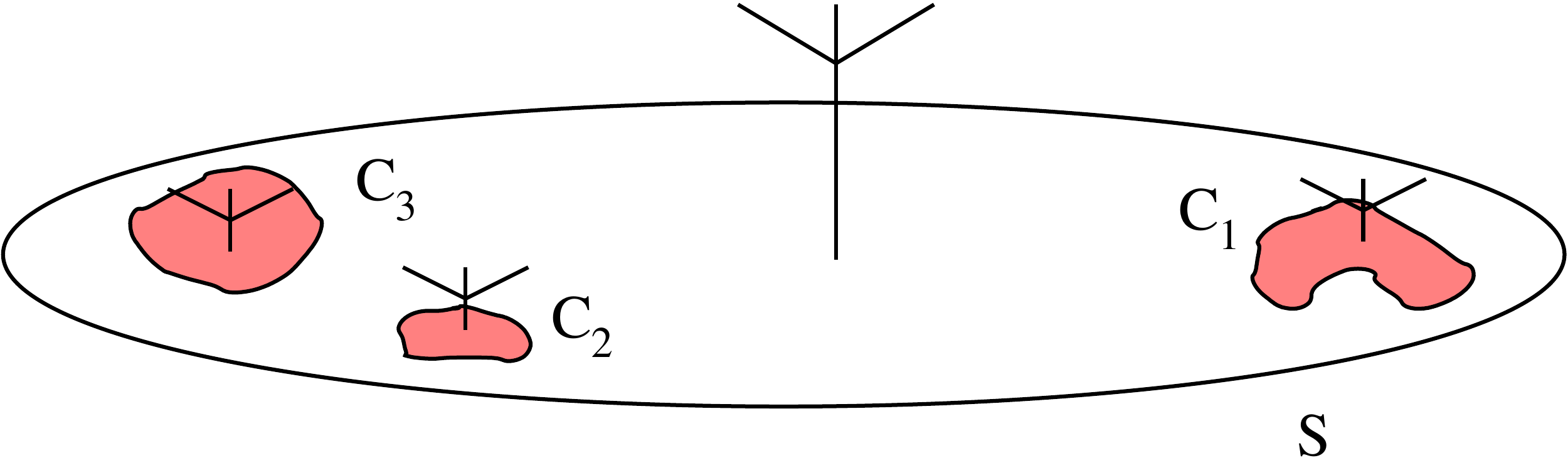}
\caption{A Macro Cell with $L=3$ Pico Cells}
\label{fig_macropico}
\end{figure}

We consider a heterogeneous network consisting of a single macro cell
containing $L \geq 1$ pico BS, see Figure \ref{fig_macropico}. $S \subset \rr^2$ denotes the (compact) coverage area of
the macro BS whilst $C_\ell \subset S, \ell=1,\cdots,L$ denote the
respective coverage areas of the pico BSs. The pico BS coverage
areas are assumed disjoint. A user at a location in $C_{\ell}$ can obtain service from
both the macro BS and the pico BS $\ell$. 
We denote the set $S - \bigcup_{l=1}^L C_{\ell}$ by $C_0$, which is the set of
locations not covered by any pico BS. Any user in $C_0$ will obtain all its support from
the macro BS.

The macro BS is assumed to use a much higher transmit power than the pico BSs, to allow it to provide
full coverage of the region $S$. We will therefore only consider scheduling policies in which the
macro time and pico times are disjoint, to avoid excessive interference at the pico BSs from the macro BS. On the other hand, pico BSs are spatially separated and use much lower power. We will therefore consider scheduling policies in which the pico BSs are allowed to operate at the same time.

A pico BS $\ell$ will use any allocated pico time to send to its users, unless there is no demand for data from within $C_{\ell}$, in
which case it switches off. The switching on and off of pico BSs has the potential to complicate the analysis, as we will see in Section~\ref{sec_genschedule}. Initially, we avoid this complication, by assuming that picocells don't interfere. In this case, the raw bit rates offered by the BSs
do not depend on the traffic in the network; they are deterministic functions of the location of the user in the network (see below). In Section~\ref{sec_genschedule}, we will allow picocells to interfere, at the expense of a more complicated model.

We suppose that users want to download files from the network, and that file requests arrive as a Poisson stream with net arrival
rate $\lambda_S$ files/sec and that the $n$th arrival 
is for a single file of (random) length $D_n < \Dbar$ bits to
be downloaded from the network. File lengths are independent and identically distributed ({\it i.i.d.}), from a common distribution $F_B$, with $\expect{D_n} = D$ bits. The locations of the arrivals are chosen
independently at random according to a continuous density $\nm{d\xi}$
with support on $S$ and bounded uniformly away from 0. Mobiles remain
fixed at their initial location until they obtain their file. Hence
in unit time, the expected number of arrivals at the vicinity of a point
$\xi$ in the macro cell coverage area $S$ is given by
$\xm{d\xi} = \lambda_S \nm{d\xi} $. 

The probability that an arrival file request is allocated to pico-region $\calC_l$ is given by
\begin{equation}
\eta_\ell = \int_{\calC_\ell} \nm{d\xi},~~\ell = 0, 1, 2, \ldots L
\end{equation}
including also the region $\calC_0$ that is served only by the macro BS.
The arrivals to each region are independent Poisson processes with rates $\lambda_S \eta_\ell~~\ell = 0, 1, 2, \ldots, L$. The conditional density in region $\calC_\ell$ is given by
\begin{equation}
\eta_\ell(d\xi) = (\nm{d\xi})/\eta_\ell
\end{equation}

The physical transmission rate of a user is determined by the user's fixed
location $\xi$. All locations are in the macro BS coverage area, and the
corresponding physical rate provided by the macro BS to location $\xi$, if scheduled, is $S_0(\xi)$ bits/sec. If the location is
within the coverage area of pico BS $\ell$, then an alternative rate is $R_\ell(\xi)$ bits/sec, provided by pico BS $\ell$, if scheduled. The macro and pico BSs are scheduled at different times, so a mobile can receive data from
both types of BS. The average data rate offered to a location will depend on the higher-layer controls: the cell association (pico or macro) and the time allocation
offered by the selected BS(s). 

We assume that rates depend continuously on location
with $0 < R_{\min} \leq R_\ell(\xi) \leq R_{\max},\ell=1,\cdots,L$ for pico cell rates
and with corresponding bounds for the macro cell rates, $S_{\min},S_{\max}$ for $S_\ell(\xi), \ell=0,\cdots,L$.
As such they are random variables with measure induced by the density $\eta$.
Define $\Rmin \doteq \min\lc R_{\min}, S_{\min} \rc > 0$ and $\Rmax$ similarly. Next
define $\rho_\ell(\xi) \doteq R_\ell(\xi)/S_\ell(\xi) > 0,\ell=1,\cdots,L$ to be the rate ratios for respective pico cell users. Clearly
$\rhoellscre \doteq \inf \lc \rho_\ell(\xi) : \xi \in C_\ell \rc > 0$ and
$\rhoellbar \doteq \sup \lc \rho_\ell(\xi) : \xi \in C_\ell \rc$. Finally we suppose that the rate-ratio
random variable has a continuous density $h,~ h_{\rho_\ell}(\rhoellscre+) > 0$ and $h_{\rho_\ell}(\rhoellbar-)>0.$


We now define the class of scheduling policies under consideration in this paper. Time is slotted with index $t \in \ZO$ where $\ZO$ is the set of non-negative integers. Scheduling decisions are made at the start of each timeslot $t$.  In particular, the first time that a file can be scheduled is in the first full time slot after its arrival. At each time slot, $t$, the scheduler chooses a fraction of the timeslot, $f_t$, that it will devote to the pico BSs, and the remaining fraction, $1 - f_t$, is allocated to the macro BS.  Once allocated this time may be divided continuously between the macro and pico users. Note that the pico BSs can all transmit in parallel, as they do not interfere. In addition, users can receive their file partly from the macro BS and partly from their pico BS, if they are within coverage. The fraction of each will be determined by the scheduling policy, to be described below. BSs transmit to one user at a time but can switch between users arbitrarily.

We will only consider non-random scheduling policies so that the scheduling outcome is determined
by the sequence of inter-arrival times of the users, together with their locations and the file sizes themselves i.e. $\Omega = \lb \rr_+,S, (0,\uB] \rb^{\nn}$
where $\rr_+$ denotes the set of non-negative real numbers.  Also we restrict to policies which clear users -
that is those schedules which transmit all user files within finite time. This is a mild assumption ensuring
all users will leave the network eventually. We term such schedules clearing schedules. Schedules such as first come first served, or processor sharing, have this property,
under Poisson arrivals with constant rate.

Within this class of schedules, are those which determine the amount of service from the macro cell and the pico cell only according to location and file size. For such schedules $\pi$,  we define
$0 \leq x^\pi_\ell(\xi,F) \leq F \leq \uB$  to be the number of bits delivered by  pico $\ell$, for a user which  arrives at point $\xi \in C_{\ell}$, with data demand $F$ bits, and the remainder is therefore delivered by the macro BS. We denote the remainder by $y^\pi_\ell(\xi,F) \doteq F - x^\pi_\ell(\xi,F)$. In the case that $\xi \in C_0$, we define $x^\pi_0(\xi,F):=0$, and $y^\pi_0(\xi,F) := F$. The network can be stabilized using only clearing policies of this type.

Given such a schedule $\pi$,  and an outcome $\omega \in \Omega$  define $N^\pi_t(\omega), t \in \ZO$, to
be the number of mobiles present at the start of slot $t$, under schedule $\pi$. We are now ready to make the following definition,
\begin{definition}
The network is said to be {\em stable} under a clearing schedule $\pi$ iff there exists $B_U > 0$ such that
\begin{equation}
\expect{N_t^\pi} < B_U,~\forall t \in \ZO
\label{eqn_stability}
\end{equation}
\end{definition}

For policies $\pi$ such that a limiting stationary distribution exists, stability implies that transmission delays have bounded first moments.

\subsection{A Continuous Linear Program}
\label{sec_ctsLP}
We begin by investigating the build up of work in the network over time.
To be specific suppose we fix a schedule $\pi$ and an outcome $\omega \in \Omega$. Then
define $V^\pi_T(\omega)$ to be the total transmission time needed to clear all users arriving in the
first $T$ slots under policy $\pi$. We now seek to construct a
location based policy $\pi$ with the property that \eqref{eqn_stability} holds, and
\begin{equation}
\lim \frac{V^\pi_T(\omega)}{T} = \tstar_\pi < 1
\label{eqn_util}
\end{equation}
almost surely. 

Consider a scenario in which there is a fixed set of users present in a HetNet with their files ready for transmission. The question arises what is the shortest time for all user files to be transmitted? The solution to this problem can be obtained via a LP as shown in \cite{ICC2013}, and which we now
present.

Let the time allocated to the pico cells (which can serve their users concurrently) be denoted by $f$ seconds.
Also, let $x_{\ell,n}$ and $y_{\ell,n}$ represent the amount of data (in bits)
received by user $n \in \calC_\ell = \{1, \dots, N_\ell\}$ in pico cell~$\ell$
from that pico cell and the macro cell, respectively. Finally let $R_{\ell,n},S_{\ell,n}$ be the corresponding
rates,
\begin{eqnarray}
\min & & f + \suml \sum_{n \in \calC_\ell} \frac{y_{\ell,n}}{S_{\ell,n}}
\label{eqn_overall_objective} \\
\mbox{sub} & & \sum_{n \in \calC_\ell} \frac{x_{\ell,n}}{R_{\ell,n}} \leq f ~~~\forall \ell \\
& & x_{\ell,n} + y_{\ell,n} \geq D_n ~~~\forall \ell, ~\forall n \in \calC_\ell
\label{eqn_overall_dataconstraint} \\
& & f \geq 0, x_{\ell,n} \geq 0, y_{\ell,n} \geq 0 ~~~\forall \ell , ~\forall n \in \calC_\ell
\label{eqn_overall_positivity_constraint}
\end{eqnarray}
Using obvious notation, the minimum value can be written $H({\bf R},{\bf S},{\bf D},{\bf L})$, where ${\bf L}$ identifies the respective mobile's picocells . It was shown in
\cite{ICC2013} that there is a set of constants
$\rho_\ell > 0, \ell=1,\cdots,L$, which determine the optimal solution. For users in pico cell $\ell$
such that
\begin{equation}
\frac{R_{\ell,n}}{S_{\ell,n}} > \rho_\ell  \label{eqn_rate_ratio_test}
\end{equation}
it is optimal to have $x_{\ell,n} := D_n$ and if the reverse inequality holds then $y_{\ell,n} := D_n$. Only where there is equality in \eqref{eqn_rate_ratio_test} can it be that both
$x_{\ell,n},y_{\ell,n}$ are positive in the optimal solution. The thresholds $\rho_\ell$ depend on the locations and demands of all the mobiles in the network.

The above LP has a continuous analogue in which $x_{\ell,n}, y_{\ell,n}$ are replaced by
$x_\ell(\xi), y_\ell(\xi) = D - x_\ell(\xi)$ where $x_\ell(\xi)$ are (Lebesgue) integrable functions.
The LP becomes,
\begin{eqnarray}
\min & & f + \sum_{\ell=0}^L \int \frac{y_\ell(\xi)}{S_\ell(\xi)} \xm{d\xi}
\label{eqn_cts_overall_objective} \\
\mbox{sub} & & \int \frac{x_\ell(\xi)}{R_\ell(\xi)} \xm{d\xi} \leq f ~~~\ell =1,\cdots,L
\label{eqn_cts_overall_dataconstraint}
\end{eqnarray}
Here, the individual mobiles are replaced by a continuous mean density and sums are
replaced by integrals. The data demands are replaced by the mean demand, $D$. The objective (\ref{eqn_cts_overall_objective}) can be interpreted as
the fraction of time that the network must be active (i.e. transmitting) in order that the traffic
as determined by $\lambda(d\xi)$ can be met. The problem itself is one in the calculus of variations,
over functions $x_\ell(\xi) \in [0,D], \ell =1,\cdots,L$ and non-zero only in $C_\ell$. $f$ can be taken
as the maximum in (\ref{eqn_cts_overall_dataconstraint}). Since the space of such functions is
compact in ${\cal L}^1$, and the map leading to the objective is continuous, the minimum is achieved.

Before we present a characterization of the solution to (\ref{eqn_cts_overall_objective})-\eqref{eqn_cts_overall_dataconstraint}, let us consider further the interpretation in terms of
a continuous mean density of demand for data from the network. If pico $\ell$ carries all the (mean) data demand from region $\calC_\ell$ then it needs time $\fellbar = \int_{\calC_\ell} \frac{D}{R_\ell(\xi)} \xm{d\xi}$. Let $\fbar = \max_{\ell=1}^L \fellbar$ be the maximum such time,
and let $l_m$ be the pico BS index that requires the maximum such time over all the pico BSs. If pico time $f$ is available, and $f > \fbar$ then the only demand on the macro BS
will come from the region $\calC_0$. Typically, however, such values of $f$ are excessive, and a smaller value in (\ref{eqn_cts_overall_objective}) can be
obtained, as in the optimal solution characterized in Theorem~\ref{thm_ctsLPsoln}.
To state the theorem define $A_{\rell} \doteq \lc \xi \in {\cal C}_\ell : \rho_l(\xi) > \rell \rc$.

\begin{theorem}
\label{thm_ctsLPsoln}
Suppose $\fstar$ is an optimal choice for 
\eqref{eqn_cts_overall_objective}-\eqref{eqn_cts_overall_dataconstraint} and that the optimum value is $\tstar$.
Then there exist $\rell \in \ls \rhoellscre, \rhoellbar \rs$ such that
{\small
\begin{equation}
\tstar = \fstar + \sum_{\ell=0}^L \int \frac{\ystar_\ell(\xi)}{S_\ell(\xi)} \xm{d\xi}
\label{eqn_tstar}
\end{equation}
}
where $\ystar_\ell(\xi) = D - \xstar_\ell(\xi) \geq 0,~\ell=1,\cdots,L$, and
{\small
\begin{eqnarray*}
x_\ell^*(\xi) & = & \left\{ \begin{array}{cc} D & \rho_l(\xi) > \rho_l^* \\ 0 & \rho_l(\xi) \leq \rho_\ell^* \end{array} \right.
\end{eqnarray*}
}
and
\begin{equation}
\fstar \geq \int_{A_{\rell}} \frac{D}{R_\ell(\xi)} \xm{d\xi},~\ell=1,\cdots,L
\label{eqn_fstarcon}
\end{equation}
\end{theorem}

{\bf Proof}: See Appendix~\ref{appendix_main_proof}.

As might be expected, the optimal solution takes the form of a rate ratio rule, as it did with the discrete LP.
Also the optimal $\xstar_\ell(\xi), \ell=1,\cdots,L$ is unique up to a set of measure 0.
This is immaterial as far as determining the capacity is concerned. Finally
Theorem~\ref{thm_ctsLPsoln} only states that suitable rate ratio thresholds can be found but does not
show how to determine them. It is desirable to obtain numerical solutions to (\ref{eqn_cts_overall_objective})-\eqref{eqn_cts_overall_dataconstraint}, so we now obtain additional results which characterise the solution further.

Given 
$f \geq 0$ and an index, $\ell$, denote by $\tau_\ell(f)$ the optimal solution to the
subproblem,
\begin{eqnarray}
\min & & \int_{\calC_\ell} \frac{\lb D - x_\ell(\xi) \rb}{S_\ell(\xi)} \xm{d\xi}
\label{eqn_subproblem_objective} \\
\mbox{sub} & & 0 \leq x_\ell(\xi) \leq D ~~\forall \xi \in \calC_\ell \label{eqn_subproblem_constraint1}\\
& & \int_{\calC_\ell} \frac{x_\ell(\xi)}{R_\ell(\xi)} \xm{d\xi} \leq f. \label{eqn_subproblem_constraint2}
\end{eqnarray}
It is readily seen that $\tau_\ell(f)$ is decreasing and convex in $f$ and that $\tau_\ell(f) = 0, f \geq \fbar_\ell$. Define
\begin{equation}
\rho_\ell(f) \doteq \sup\{x: \int_{A_{x}} \frac{D}{R_\ell(\xi)} \xm{d\xi}  > f \}.
\label{eqn_defrho}
\end{equation}
for $f \in [0,\fbar_\ell]$, where $A_x \subseteq {\cal C}_\ell$ is understood. By definition
$\rell(0) = \rhoellbar$ and $\rell(\fbar_\ell) = \rhoellscre$. By continuity of the integral,
$$
\int_{A_{\rho_\ell(f)}} \frac{D}{R_\ell(\xi)} \xm{d \xi} = f,
$$
and, moreover, $\rho_\ell(f)$ is strictly decreasing, although not necessarily continuous.

Given $f \in (0, \fbar_\ell]$ for suitable $\deltf > 0$,
\begin{eqnarray}
\tau_\ell(f) - \tau_\ell(f - \deltf) & = & - \int_{A^c_{\rell(f)}-A^c_{\rell(f -\deltf)}} \frac{D}{S_\ell(\xi)} \xm{d\xi} \nonumber\\ 
& = &- \trho \int_{A_{\rell(f-\deltf)-A_{\rell(f)}} } \frac{D}{R_\ell(\xi)} \xm{d\xi} =  -\trho \deltf,
\end{eqnarray}
where $\trho \in \ls \rell(f) , \rell(f - \deltf) \rs$, by the mean value theorem. It follows that
the left derivative satisfies $\ls \tau_\ell(f) \rs^{'}_{-} \leq - \rell(f)$ and similarly for the right
derivative, $\ls \tau_\ell(f) \rs^{'}_{+} \geq - \rell(f)$.

We thus see that $\tau_\ell(f)$ is strictly convex in $f$ over the interval $[0,\fbar_\ell]$ because $\rho_\ell(f)$
is strictly decreasing. Hence so is the
function $\tau(f) := f + \sum_{\ell=0}^L \tau_\ell(f)$ which therefore has a unique optimum at $\fstar$.
It follows that the edge condition
\begin{equation}
1 \geq \sum_{\ell} \rho_\ell(\fstar)
\label{eqn_edge}
\end{equation}
must hold and with the reverse inequality on replacing $\fstar$ with $f < \fstar$. Equality holds in \eqref{eqn_edge} if $\tau$
is differentiable at $\fstar$. It follows that in case $\sum_\ell \rhoellbar  \leq 1$, $\fstar = 0$ (no pico time) is optimal and in case $\sum_\ell \rhoellscre  \geq 1$, $\fstar = \fbar$ (all pico time) is optimal. Otherwise the solution
satisfies $\fstar \in (0, \fbar)$.

Numerical results can be obtained by first checking the edge condition (\ref{eqn_edge}) for extreme choices
of $\rho_\ell$. In case $\sum_\ell \rhoellbar  > 1 > \sum_\ell \rhoellscre$ the optimum may be
found by determining $\tau_\ell(f)$. This in turn can be evaluated once $\rho_\ell(f)$ is obtained, which
can be done by a one dimensional search and numerical integration.

\section{Capacity Results}
\label{sec_capres}
\subsection{The Achievability Result}
\label{sec_achievable}
In this section, we assume that the solution, $\tau^*$, to \eqref{eqn_cts_overall_objective}-\eqref{eqn_cts_overall_dataconstraint} satisfies $\tau^* < 1$. We will now propose a scheduling policy that we will demonstrate is stable. In fact, the discussion in Section \ref{sec_ctsLP} suggests the following approach to scheduling: A rate ratio policy is defined via
a vector ${\bf a} \in \R^L, a_\ell \in \rhoint$. It works by assigning pico time according to ${\bf a}$, that is mobiles in
pico cell $\ell$ at location $\xi$ are transmitted only using pico time if
$\al < \rho_l(\xi)$,
all other mobiles are transmitted using only macro cell time. In the following we choose
$a_\ell := \rho_\ell^*, \ell = 0, \cdots, L$. To each timeslot allocate a fraction
$\fstar/\tstar$ to the pico cells and the remaining time to the macro cell. We will suppose that time is
divided continuously during the slot so that $u$ seconds into the slot $u \fstar/\tstar$ has been allocated to
the picos.

Files are transmitted 
at a rate which is reduced by the time sharing fractions as stated above. This is for all $L+1$ servers, the $L$ picos and the macro cell. For the purposes
of analysis we further suppose that the multiple mobiles, which can arrive between one slot and the next,
have their files merged in order of arrival, to be counted as a single job to be served. We assume that the server allocates its time to these jobs in first come first served (FCFS) order.
Clearly the sequence of merged job service
times $X^\ell_t$ (in seconds) at the start of each slot are independent and identically distributed ({{\it i.i.d.}) by construction. There is a positive probability of a job requiring zero time (corresponding to no arrivals of files during the previous slot).

There is one job arrival at the beginning of each time slot, and the service time is generally distributed, hence each queue is D/G/1. To compute the mean job duration, we note that each individual file has a random length, with average value $D$ bits. The number of arrivals to the pico-cell $\ell$ queue in a slot is Poisson with mean $\lambda_S \nu_l T_S$, where $T_S$ is the length of the slot (in seconds) and $\nu_l$ is the probability of the file being allocated to pico BS $\ell$, namely
{\small
\begin{equation}
\nu_\ell = \int_{\calC_\ell} I[\rho_\ell(\xi) > \rho_\ell^*] \nm{d\xi}
\end{equation}
}
The average time each individual file requires from the pico BS $\ell$ is
{\small
\begin{equation}
\FellP = \nu_l^{-1} \left(\int_{\lc \xi \in {\cal C}_\ell : \rho_\ell(\xi) > \rho_\ell^* \rc} \frac{D}{R_\ell(\xi)} \nm{d \xi}\right)
\end{equation}
}
Thus, the 
workload, measured in seconds of work per second is,
{\small
\begin{equation}
\lambda_S \FellP = \int_{\lc \xi \in {\cal C}_\ell : \rho_\ell(\xi) > \rho_\ell^* \rc} \frac{D}{R_\ell(\xi)} \xm{d \xi} \leq \fstar
\label{eqn_expected_service_pico}
\end{equation}
}
the last inequality from \eqref{eqn_fstarcon}. The service rate is $f^*/\tau^*$ seconds per second, so the utilization is at most $\tau^*$.

The number of arrivals to the macro-cell queue in a slot is Poisson with mean $\lambda_S (1 - \sum_{\ell=1}^L \nu_\ell) T_s$. The average time each individual file requires from the pico BS $\ell$ is
{\small
\begin{equation}
\FM = (1 - \sum_{\ell=1}^L \nu_\ell)^{-1} (\FzeroM + \sum_{\ell=1}^L \FellM)
\end{equation}
}
where
{\small
\begin{eqnarray}
\FzeroM & = & \int_{\calC_0} \frac{D}{S_0(\xi)} \nm{d\xi} \\
\FellM & = & \int_{\calC_\ell} \frac{D}{S_\ell(\xi)} I[\rho_\ell(\xi) < \rho_\ell^*] ~\nm{d\xi}.
\end{eqnarray}
}
Thus, the workload at the macro-cell queue, measured in seconds of work per second, is
{\small
\begin{equation}
\lambda_S \FM = \int_{\calC_0} \frac{D}{S_0(\xi)} \xm{d\xi}  + \sum_{\ell=1}^L \int_{\lc \xi \in {\cal C}_\ell : \rho_\ell(\xi) < \rho_\ell^* \rc} \frac{D}{R_\ell(\xi)} \xm{d \xi}
\label{eqn_expected_service_macro}
\end{equation}
}
But the RHS of \eqref{eqn_expected_service_macro} is no more than $\tau^* - f^*$, by \eqref{eqn_tstar}. The service rate is $(\tau^* - f^*)/\tau^*$ seconds per second, so the utilization of the macro-cell queue is $\tau^*$. 

As far as delay is concerned, this can be broken down to i) time before entering queue, ii) queueing time, iii)
time to clear earlier files in the merged job. The mean delay for i) is at most $T_S$ and for iii)
no more than $\lambda_S \nu_\ell T_S \times (\Dbar/\Rmin)$. As far as ii) is concerned we have just shown the utilisation for picos and macros is no more than $\tstar$.
The well known theory of the $D/G/1$ queue, see \cite{Spitzer} shows that the sequence of waiting times $W_n$ has a limiting distribution with finite expectation provided that the sum on the RHS of
{\small
\begin{equation}
\expect{W_\infty} = \sum_{n=1}^\infty \frac{1}{n} \expect{S_n^+}
\label{eqn_Spitzer}
\end{equation}
}
is finite, where $S_n \doteq \sum_{k=1}^n X^\ell_k - n T_S$. This is
readily shown to be the case. This is because the mean centred version of $X^\ell_1 - T_S$ has
finite moments as a consequence of our assumptions. 

From the above we may deduce the following theorem,
\begin{theorem}
\label{theorem_fwd}
If the solution to the continuous LP, satisfies $\tstar < 1$ determined via (\ref{eqn_cts_overall_objective})-\eqref{eqn_cts_overall_dataconstraint}
then there is a policy $\pi$  and a constant $U > 0$ such that the expected number
$N_n$ of mobiles still in the network at slot $n$ satisfies
\begin{equation}
  {\mathbb E}_{\pi} \lb N_n \rb < U,~\forall n \in \NN
\end{equation}
and $\pi$ is the time sharing scheduler as determined by $\fstar$ and $\rho_\ell^*, \ell =1 ,\cdots,L$.
\end{theorem}

Of course, scheduling file downloads in a FCFS fashion as described above, is far from practical. A more
practical scheme is to transmit files simultaneously by sharing the available bandwidth evenly amongst
the waiting users. Moreover since the ABS slots $T_S$ are very small compared with times to transmit
files we can treat these times as infinitesimal and model the transmission delays using the
M/G/1/PS model; see Section~\ref{sec-PS}. This model will provide representative, if not ideal, performance results for actual networks and results are presented in Section \ref{sec_numer}.

\subsection{Converse}
\label{sec_converse}

In this section we will show that if $\tau^*$, the infimum of the solutions to (\ref{eqn_cts_overall_objective}), is strictly greater than 1
then for any clearing schedule $\pi$ the total residual transmission time
increases to infinity. To be more specific, given a scheduling policy $\pi$
let $\VTp$ be the (random) total time that transmission of any of the
arrivals during $[0,T]$ is taking place under schedule $\pi$. In other
words, the total time the network is actively transmitting the file of at least
one of the users which arrived in $[0,T]$. This time is not assumed to be accrued continuously: There can be interspersed services of later arrivals, but we do not count the
service of the later arrivals in $\VTp$.

For fixed $T$ and a given
set of user arrivals, $\VTp$ is upper bounded by the time which would be
taken if user files are transmitted one after the other at rate $\Rmin$.
It is lower bounded by the time taken if the files are transmitted sequentially
at rate $(L+1)\Rmax$.

We now state the converse theorem.
\begin{theorem}
\label{thm_main}
If $\tau^* > 1$ then there is a constant $\eta > 0$ such that for any clearing schedule $\pi$
it holds that
\begin{equation}
\liminf_{T} \frac{\VTp(\omega)}{T} > 1 + \eta
\end{equation}
almost surely.
\end{theorem}

Hence after time $T$ the residual work (time to clear
the remaining users) is at least $\eta T$ for all $T$ sufficiently large. The implication is that with probability 1 and for all $T$
sufficiently large there must be at least
\begin{equation}
\NTp(\omega) \geq \floor{\frac{\eta T \Rmin}{\uB}}
\label{eqn_Ntogo}
\end{equation}
users present at time $T$. This is because no clearing schedule transmits files at a rate smaller than $\Rmin$ at any time. Thus any clearing schedule must have at least $\eta T \Rmin$ bits to be transmitted and as there are at most $\uB$ bits per file there must be at least as many users still in the network as expressed in the RHS of (\ref{eqn_Ntogo}).


The idea of the proof is that first, for any realisation over
$T$ slots we can never use less time than the solution to the corresponding discrete
LP, see \eqref{eqn_overall_objective}. Second it will be shown that if the optimal
solution to (\ref{eqn_cts_overall_objective}) is
$$
\tstar = \fstar + \sum_{\ell=1}^L \int_S \frac{\ystar_\ell}{S_\xi} \lambda(d\xi)
 > 1
$$
then there is a constant $\eta > 0$ such that
\begin{equation}
\liminf_T \frac{\VLP(\omega)}{T} > 1 + \frac{\eta}{2} > 1
\label{eqn_LPbuildsup}
\end{equation}
almost surely. This implies that the residual amount of work grows
linearly over time. \\
{\bf Proof of Theorem \ref{thm_main}} \\
Assume that $\tau^* > 1$, and let $N_T$ be the number of file requests which arrive in the interval $[0,T]$.
Set ${\bf R} := \lb R(\xi_n) \rb_{n=1}^{N_T}$ and ${\bf S},{\bf D},{\bf L}$ being set similarly.
Define $\VLP(\omega) \doteq H({\bf R},{\bf S},{\bf D},{\bf L})$ which is clearly a well defined random variable.
The optimal solution to the LP is characterized by rate-ratio thresholds, $\rho_{1,T}^*, \rho_{2,T}^*, \ldots \rho_{L,T}^*$, and which are random variables (proof omitted).

$\VLP$ lower bounds the time needed to clear the users who arrive in $[0,T]$ under any clearing schedule:
\begin{lemma}
\label{lem_sample}
For all sample paths $\omega$ and for any clearing schedule $\pi$,
\begin{equation}
\liminf_T \frac{\VLP(\omega)}{T} \leq \liminf_T \frac{\VTp(\omega)}{T}
\label{eqn_liminf}
\end{equation}
\end{lemma}
{\bf Proof} \\
Recall that $N_T$ is the number of file requests which arrive in the interval $[0,T]$ and we have indexed the users in order of arrival,
$n=1,\cdots,N_T$. Let $x_n,y_n$ denote the number of bits sent by the $n$th users pico cell (if any) and the
bits sent by the macro cell under clearing policy $\pi$. Since $\pi$ is a clearing schedule,  it must be the case that,
\begin{equation}
x_n + y_n \geq D_n, \forall n=1,\cdots,N_T
\label{eqn_filepi}
\end{equation}
Let $\fpT$ be the total amount of
time during which at least one of the $N_T$ mobiles is getting its file from the pico BS.
It follows that,
\begin{equation}
\fpT \geq \sum_{n: \xi_n \in C_\ell} \frac{x_n}{R_n}
\label{eqn_picopi}
\end{equation}
for each $\ell=1,\cdots,L$. Finally by definition,
\begin{equation}
\fpT + \sum_{n=1}^{N_T} \frac{y_n}{S_n} = \VTp(\omega)
\label{eqn_VTpidefn}
\end{equation}
From (\ref{eqn_filepi}),(\ref{eqn_picopi}) it can be seen that the LP constraints \eqref{eqn_overall_dataconstraint}-\eqref{eqn_overall_positivity_constraint} are satisfied. Therefore
$\VLP(\omega) \leq \VTp(\omega)$, for any sample path.  We thus obtain
(\ref{eqn_liminf}) which completes the proof. \qed.

To proceed, consider an alternative scenario in which files, instead of arriving according to a Poisson process during the interval $(0,T)$, are all present in the system from time zero. Subsequent files are discarded. $\VLP$ is then the minimum amount of time required to service these users. In this LP assignment, the pico BSs are allocated time $f_T^*$, and the macro is allocated time $\VLP - f_T^*$, and the users are assigned according to the rate-ratio thresholds as obtained from the solution to \eqref{eqn_overall_objective}. We would like to show \eqref{eqn_LPbuildsup} directly, but it is easier to first analyze a simpler, static approach.

Consider a simpler static system in which fixed rate-ratio thresholds are used for the assignment of user to BS. To this end, let $\rho^{(a)} = (\rho^{(a)}_1, \rho^{(a)}_2, \ldots, \rho^{(a)}_L)$ be a fixed vector of rate-ratio thresholds. Define
\begin{eqnarray}
f_{\ell,T}^{(a)} & \doteq & \sum_{n: \xi_n \in \calC_\ell} \frac{D_n}{R_n} I[\rho_n > \rho_\ell^{(a)}] 
\end{eqnarray}
and set $f_T^{(a)} \doteq \maxl f_{\ell,T}^{(a)}$. Additionally define,
\begin{equation}
Y_T^{(a)} \doteq \sum_{n: \xi_n \in \calC_0} \frac{D_n}{S_n} + \suml \sum_{n: \xi_n \in \calC_\ell} \frac{D_n}{S_n} I[\rho_n < \rho_\ell^{(a)}] 
\end{equation}
Lemma~\ref{lem_LLN} below shows that the law of large numbers applies to these quantities, with the following deterministic limits
\begin{equation}
f_{\ell}^{(a)} \doteq  \int_{A_{\rho_\ell^{(a)}}} \frac{D-y_\ell^{(a)}}{R_\ell(\xi)} \xm{d\xi}
\end{equation}
and $f^{(a)} \doteq \maxl f_{\ell}^{(a)}$. Also define,
\begin{equation}
Y^{(a)} \doteq \int_{\calC_0} \frac{y_\ell(\xi)}{S_\ell(\xi)} + \suml \int_{A^c_{\rho_\ell^{(a)}}} \frac{y_\ell(\xi)}{S_\ell(\xi)} \xm{d\xi}
\end{equation}
where
\begin{equation}
y_\ell(\xi) = \left\{ \begin{array}{cc} 0 & \rho_\ell(\xi) > \rho_\ell^{(a)} \\ D & \mbox{~o.w.~} \end{array} \right. ~~\xi \in \calC_\ell~~\ell=1, 2, \ldots, L
\label{eqn_yell_rhoa}
\end{equation}

\begin{lemma}
\label{lem_LLN}
The following limits hold almost surely, 
\begin{eqnarray}
\lim_{T \uparrow \infty} \frac{1}{T} f_T^{(a)}(\omega) & = & f^{(a)} \\
\lim_{T \uparrow \infty} \frac{1}{T} Y_T^{(a)}(\omega) & = & Y^{(a)}
\end{eqnarray}
\end{lemma}
{\bf Proof}: See Appendix~\ref{app_lem_LLN}.

Define  $V^{(\bfa)}_T(\omega) :=  f_T^{(a)} + Y_T^{(a)}$ to be the total time used. Since the $y_\ell(\cdot)$ in \eqref{eqn_yell_rhoa} (and the corresponding $x_\ell(\cdot)$) are feasible for \eqref{eqn_cts_overall_objective}-\eqref{eqn_cts_overall_dataconstraint},  it follows from the above lemma
that there exists $\eta > 0$ such that 
\begin{equation}
\lim_T \frac{V^{(\bfa)}_T}{T} \geq 1 + \eta~~a.s. 
\label{eqn_Vaeta}
\end{equation}

Any vector ${\bf \rho^{(a)}}$ can be considered as an assignment rule, since it indicates to which BS the arrival will go for service. We now define a discrete set ${\mathcal A}$ of such assignment rules. For a fixed integer $N_P$, to be taken large, define the constants $(\rho_\ell(i))_{i=1}^{N_P+1}$, $\ell = 1, 2, \ldots L$, by
\begin{equation}
\rho_\ell(i) = \left(1 - \frac{i-1}{N_P}\right) \rhoellscre + \left(\frac{i-1}{N_P}\right) \rhoellbar
\label{eqn_def_points}
\end{equation}
and define
\begin{eqnarray}
{\mathcal A}_\ell & = & \{\rho_\ell(i): i = 1, 2, \ldots, N_P\} \\
H_{\ell,i} & = & (\rho_\ell(i), \rho_\ell(i+1))~~i = 1, 2, \ldots N_P
\end{eqnarray}
so the interval $(\rhoellscre,\rhoellbar)$ is divided into $N_P$ equal subintervals, $(H_{\ell,i})_{i=1}^{N_P}$, and ${\mathcal A}_\ell$ is the set of left endpoints of those subintervals. Define ${\mathcal A}$ to be the set of $N_P^L$ vectors given by
\begin{equation}
{\mathcal A} = \prod_{\ell=1}^L {\mathcal A}_\ell
\end{equation}

The next step is to approximate the rate-ratio vector selected by the LP assignment, $\rho_T^*$, with a vector that is close to it in the set ${\mathcal A}$. For each $\ell$, let $H_{\ell,n_\ell}$ be the interval containing $\rho_T^*$, with corresponding left endpoint $\rho_\ell(n_\ell)$. Define the vector ${\bf a}_T \in {\mathcal A}$ by
\begin{equation}
a_{\ell,T} = \rho_\ell(n_l)~~\ell = 1,2, \ldots, L
\end{equation}
Note that ${\bf a}_T$ is a random vector, taking values in the set ${\mathcal A}$.

Now let ${\bf a}_T$ be the assignment rule. This is the same as under the LP allocation, except for users in the interval in which the LP rate-ratio threshold lies. Lemma~\ref{lem_LDR} shows that for sufficiently large $T$, the number of users in each of these intervals is upper bounded by $L \epsilon T$.

\begin{lemma}
\label{lem_LDR}
Let $N^{n,l}_{H,T}$ be the number of arrivals into ${\calC}_\ell$ during in slots $ t \leq T$
whose rate ratio falls into $H_{\ell,n}$. Then for any $\epsilon > 0$ 
there is a fixed set of intervals, with $N_P$ sufficiently large, and a corresponding random variable $T_E < \infty$, such that for $T \geq T_E$, the following holds, for all $n$ and $l$:
$$\frac{N^{n,l}_{H,T}}{T} < \epsilon   \mbox{~~a.s.}$$
\end{lemma}
{\bf Proof} The proof of Lemma \ref{lem_LDR} is given in Appendix \ref{app_lemLDR}. \qed

Since the minimum rate is $\Rmin$, the time to service the users in each of these intervals is upper bounded by $\displaystyle \frac{L \epsilon T \uB}{\Rmin}$. It follows that for $T > T_E$,
{\small
\begin{eqnarray}
\frac{\VLP}{T} & \geq & \frac{\ATLP}{T} - \frac{L \epsilon \uB}{\Rmin} \\
& \geq & \inf_{a \in {\mathcal A}} \frac{V^a}{T} - \frac{L \epsilon \uB}{\Rmin}
\end{eqnarray}
}
Thus,
{\small
\begin{eqnarray}
\lim \inf_T \frac{\VLP}{T} & \geq & \lim \inf_T \inf_{a \in {\mathcal A}} \frac{V^a}{T} - \frac{L \epsilon \uB}{\Rmin} \\
& = & \inf_{a \in {\mathcal A}} \lim \inf_T \frac{V^a}{T} - \frac{L \epsilon \uB}{\Rmin} \\
& \geq & (1 + \eta) - \frac{L \epsilon \uB}{\Rmin}
\end{eqnarray}}
the last inequality from \eqref{eqn_Vaeta}. 
Thus, if we take $\epsilon = \frac{\eta \Rmin}{2 L \uB}$, we obtain
{\small
\begin{equation}
\lim \inf_T \frac{\VLP}{T} \geq 1 + \frac{\eta}{2}.
\end{equation}}
The proof of the Theorem concludes by noting that
{\small
\begin{equation}
\lim \inf_T \frac{\VTp}{T} \geq 1 + \frac{\eta}{2}.
\end{equation}}
holds for {\it any} clearing schedule $\pi$, due to Lemma~\ref{lem_sample}, and hence the workload must build linearly over time for any clearing schedule.
\qed

\section{Delays under Processor Sharing}
\label{sec-PS}
The proof of Theorem~\ref{theorem_fwd} only requires the consideration of a simple scheduler based on first come, first served (FCFS) processing of jobs in the queues. Theorem~\ref{thm_main} shows that this scheduler is good enough, as far as achieving stability is concerned. However, it is well known that Processor Sharing will provide much better delay performance than FCFS. Since we are only interested in the stable case in this section, we assume that $\tau^* < 1$.

In Processor Sharing, all customers in any queue (macro or pico cell) get service from the BS at all times; the service rate is split equally amongst all customers in the queue. Such systems are examples of the symmetric queues considered in Section 3.3 of Kelly~\cite{Kelly79}; in particular, they are server-sharing queues. In \cite{Kelly79} the arrivals to each queue are independent Poisson processes. We obtain Processor Sharing in the limit as we reduce the slot duration to zero, and allocate an equal time share between the jobs (files) in the system. The arrival processes are independent Poisson processes in the limit as the slot duration goes to zero.

We ignore queues which are underloaded, and which typically have much smaller delays, and instead focus on the ``bottleneck'' queues, which are the ones for which there is equality in \eqref{eqn_fstarcon} and $\rho_\ell > 0$. The macro cell queue is also a bottleneck queue. The utilization of the bottleneck servers does not depend on the service discipline, and their utilization is $\tau^*$.

It follows from Theorem~3.10 in \cite{Kelly79} that the number of customers queued in each bottleneck queue is Geometric, with parameter $\tau^*$, and this result is insensitive to the distribution of the service time of each customer, depending only on the mean service time. It follows that each bottleneck queue size is Geometric with mean $\Nbar = \frac{1}{1 - \tau^*}$. The mean delay in queue $\ell$ can be computed from Little's law, and is given by
\begin{equation}
\Tbar = (\lambda_S \eta_\ell (1 - \tau^*))^{-1}
\label{eqn_delayPS}
\end{equation}
where $\ell$ is any index in $\{0, 1, 2, \ldots L\}$ corresponding to a bottleneck queue. It can be seen that the delays in each bottleneck queue are not the same, unless the arrival rates are the same. In practice, average delays could be reduced further by equalizing delays rather than equalizing utilization.

Lower delays can be achieved if the time-share between macro and picos is allowed to adapt to the traffic. Even better, the rate-ratio thresholds can be adaptive to the traffic, instead of being fixed constants. Our results show that such approaches cannot increase capacity, but one can expect the delay performance to be improved, possibly dramatically in some cases.

\section{On-off scheduling}
\label{sec_genschedule}
Our results heretofore are based on the assumption that  the physical bit rates, $R_\ell(\xi),S_\ell(\xi)$, depend only on the location, $\xi$, of the mobile in the network. This assumption is somewhat restrictive and so in this section we
show how the notion of capacity extends to the case where the physical rates are dependent on network state.
(Note, however, that our prior assumptions are valid for networks with regions $\calC_\ell$ which are sufficiently geographically separated so that between pico cell interference is negligible and are also valid in a network with non-negligible inter-pico interference, but in which the pico BSs transmit with constant power and never go silent. These determine lower/upper bounds for capacity and upper bounds for delay in arbitrary networks which may be adequate for some purposes.)

In what follows we will restrict to the assumption that rates at a particular location depend only on the on/off state of
the picos (these are all off if the macro is on). Hence define $\Xi$ to be
be all subsets of $\lc 1,\cdots,L \rc$ with at least two members. Clearly the empty set is redundant, and the singleton sets
can be subsumed as macro rates by redefining the macro rate of a mobile to be the maximum of its macro rate and its
pico rate when scheduled by itself. Hence if $L=3$ the network states are $\lc 1,2\rc,\lc 1,3 \rc, \lc 2,3\rc,\lc 1,2,3 \rc$.
Denote a given on-off state by $\sigma$, then for each pico cell $\ell \in \sigma$ and each location there is a corresponding
pico rate $R^\sigma_\ell(\xi)$. ABS times are now periods devoted to the various
network states and we may suppose that a period $f^\sigma$ is made available for operation of
the network in state $\sigma$. 

We may now revisit the set up that we had earlier in much the same way as before. First
the discrete LP for a given set of users file lengths etc. can be solved along similar
lines to the one presented earlier. We leave the reader to check this is the case.
Second a continuous version of the LP can be obtained by replacing summation with
integration over a non-homogeneous Poisson mean measure, as was done earlier. At this point,
two issues arise. First, are there corresponding multipliers generalizing the $\rho_\ell$ used
earlier? and second, does the optimal solution determine capacity as it did before? The answer to both
questions is yes, and can be demonstrated along similar lines to those detailed earlier. We therefore confine ourselves to setting up the LP, and identifying the optimal structure.

To set up the continuous LP define, $x^\sigma_\ell(\xi)$ to be the number of bits transmitted to mobiles
at location $\xi$ in cell $\ell$, when the pico on state is $\sigma$. Let $\fsgma$ denote the fraction of
time per unit slot allocated to on state $\sigma$. Also let $f = \sum_{\sigma: L-1 \geq \abs{\sigma} \geq 2} \fsgma$ and define $y_\ell(\xi)$ to be the bits transmitted by the macro cell for mobiles in pico cell $\ell$ so that
$y_\ell(\xi) + \sum_{\sigma: \ell \in \sigma} x^\sigma_\ell(\xi) = D$ where $D$ is mean file size as before. The continuous LP is,
\begin{eqnarray}
\min & & f + \sum_{\ell=0}^L \int \frac{y_\ell(\xi)}{S_\ell(\xi)} \xm{d\xi}
\label{eqn_genctsobj} \\
\mbox{sub} & & \int \frac{x^\sigma_\ell(\xi)}{\Rsellxi} \xm{d\xi} \leq f^\sigma,~~~\ell \in \sigma
\label{eqn_ctscon}
\end{eqnarray}

We now characterize the optimal solution for (\ref{eqn_genctsobj}) in the following theorem.
\begin{theorem}
\label{thm_ctsLPsoln_modes}
Let $\bartau$ denote the optimal value, and $\barfsgma$ the optimizer of (\ref{eqn_genctsobj}). Then there
are Lagrange multipliers $\esell \geq 0, \forall \sigma \in \Xi ,~\ell \in \sigma$ such
that the optimal $\barxsell(\xi)$ satisfies
$$
 \barxsell(\xi) = \left\{
              \begin{array}{cc}
               D & \xi \in \Aseta \\
               0 & \mbox{~otherwise}
               \end{array} \right.
$$
where $\Aseta \doteq \lc \xi : \max_{\varrho} \rhorhoell(\xi)/\erell = \rhosgmaell(\xi)/\esell > 1 \rc $
and if $\barfsgma > 0$ then
$$
\sum_{\ell \in \sigma} \esell = 1.
$$
Moreover, for all $\sigma$ such that $\ell \in \sigma$ and $f^\sigma > 0$, then \eqref{eqn_ctscon} holds with equality.
\end{theorem}
{\bf Proof}: See Appendix \ref{app_ctsLPsoln_modes}.

It is helpful to interpret the form of the solution in the above theorem.  First the Lagrange multipliers $\esell$ can be interpreted as the rate of exchange of macro time for time used when operating in state $\sigma$. Since
necessarily $\esell \leq 1$  these quantities can thus be thought of as a reuse gain. Also it is
the rate ratios with the macro cell which are in the numerator and if the maximum over all states is
smaller than 1, the mobile is assigned to the macro cell (although this may be by construction actually
time with just the given pico cell on.

The final statement in Theorem~\ref{thm_ctsLPsoln_modes} means that that all picos in all modes saturate together, as capacity is approached.

\section{Numerical Results}

\label{sec_numer}
We consider a circular macrocell, with three pico BSs ($L=3$), as illustrated
in Figure~\ref{fig_scenario}. The macrocell has radius $1$ km, and the hotspots (pico regions), ${\calC}_1, {\calC}_2, {\calC}_3$ each have radius $150$ m. The macro BS is located at the origin,
and the three pico BSs are located at coordinates $\lb  (-339,741), (218,-230), (561,-457) \rb$ , respectively, with distances measured in metres.

\begin{figure}
\centering
\includegraphics[width=2.5in]{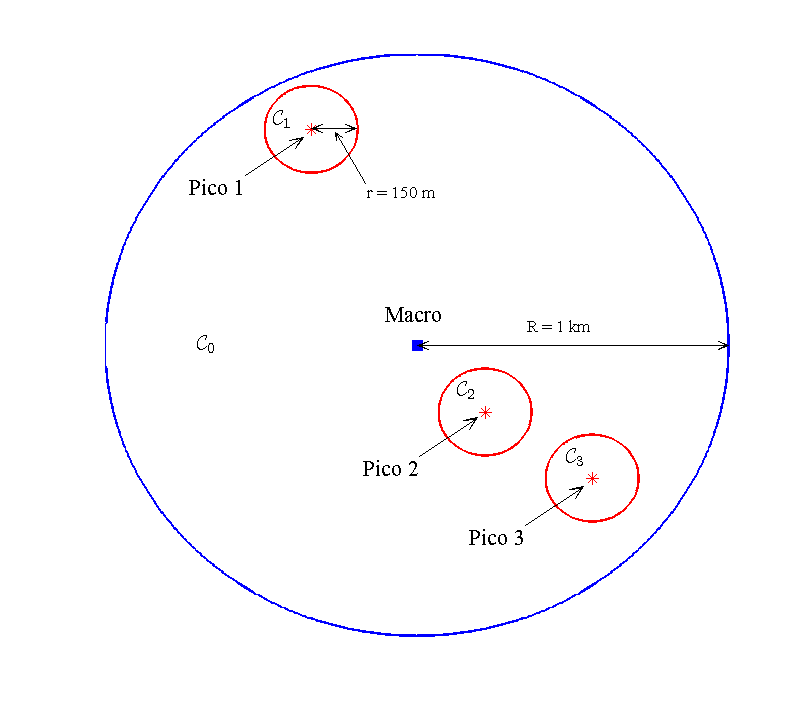}
\caption{Macro-cell containing $L=3$ Pico Cells}
\label{fig_scenario}
\end{figure}

File requests arrive as a Poisson process of rate $\lambda_S$ arrivals/sec, and they go in hotspots ${\calC}_1, {\calC}_2, {\calC}_3$ with probabilities ${\bf p} = \lb 0.4, 0.25, 0.15 \rb$, respectively. With probability $0.2$ the arrival selects the macro-only region ${\cal C}_0$. File requests falling in a hot spot are distributed uniformly in an
annulus outer radius $150~m.$ and inner radius $10~m.$ File requests falling in ${\calC}_0$ are distributed uniformly in the allowable region, which excludes the pico-regions, and excludes the area within $10$~m of the macro BS. The mean file size to be transferred was fixed at
 $D = 4Mbits$.

The macro BS transmits with power $46~dBm$ with an additional antenna gain of $14$ dBi. The macro BS to user pathloss model is
$\text{PL}(d) = 15.3 + 37.6\log_{10}(d)$ dB, where $d$ is in metres from the macro BS. The pico BS transmits with power $30~dBm$ with an additional antenna gain of $5$ dBi. The pico BS to user pathloss model is $\text{PL}(d) = 30.6 + 36.7\log_{10}(d)$ dB, where $d$ is in metres from the pico BS.  The receiver noise at the mobiles is assumed to be $-104$ dBm. This model specifies the SNR that can be achieved at any location in the macrocell, from any BS.

We consider two interference cases below. In the no-interference scenario, we base the achieveable rate on Shannon's formula for the AWGN channel, $C = W \log_1(1 + \mbox{SNR})$ bits/sec. In the alternative, we take into account interference from other pico BSs, when a mobile is receiving from a pico BS. In this case, the SINR is calculated, using the given pico BS path-loss model for the $L-1$ interfering signals, and then using $C = W \log_2(1 + \mbox{SINR})$ bits/sec to calculate the achievable rate in bits/sec. Interference only applies for pico links, the SNR is used for the macro links in both scenarios. The bandwidth, $W$ is taken to be $1$ MHz.

In the following results, all relevant integrals were estimated using Monte-Carlo simulation.

Figure \ref{fig_rhof} plots the three rate-ratio threshold functions, $\rho_\ell(f)$, described in \eqref{eqn_defrho}, for the no interference scenario. The figure illustrates the fact that these functions are decreasing. Figure~\ref{fig_tau} illustrates the fact that the minimum of $\tau(f)$ coincides with the unique solution to the edge-rate condition \eqref{eqn_edge}. In these scenarios, the function $\tau(f)$ is differentiable, so there is equality in~\eqref{eqn_edge}.

 \begin{figure}[t]
    \centering
    \begin{floatrow}
      \ffigbox[\FBwidth]{\caption{Fraction ABS Time $f$ versus Rate Ratio $\rho$ $L=3$ Pico Cells}\label{fig_rhof}}{%
        \includegraphics[width=3in]{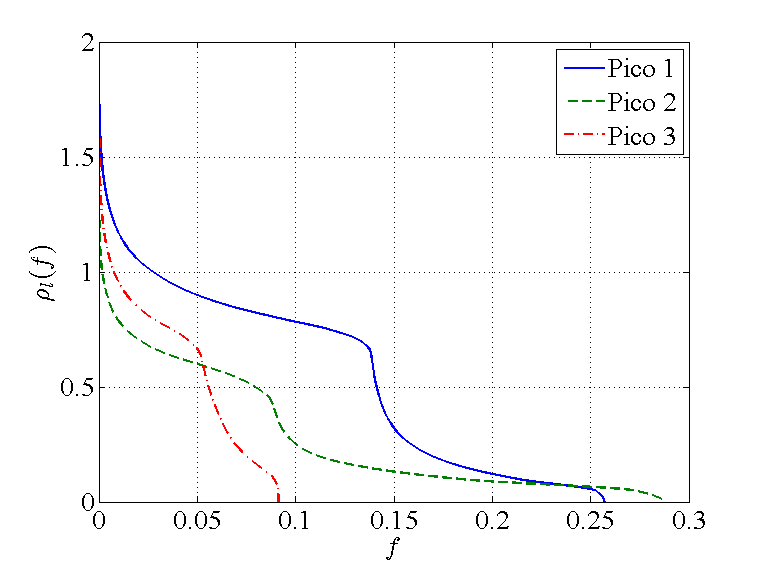}
      }
      \ffigbox[\FBwidth]{\caption{Total time $\tau$ and Edge Condition Trade off as a Function of $f$}\label{fig_tau}}{%
        \includegraphics[width=3in]{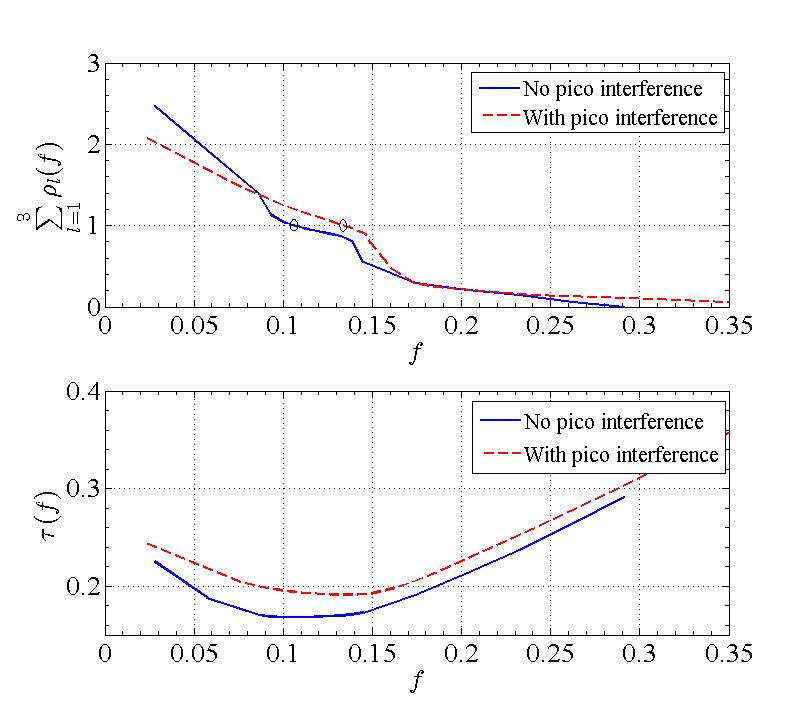}
      }
    \end{floatrow}
  \end{figure}

We should expect that the case with interference should give a more accurate account of the real system, but since the pico cells are well separated geographically, we also provide results excluding inter-pico interference to see if this interference has significant impact. As discussed earlier, no-interference is a ``best-case'' assumption and  the interference assumption is actually ``worst-case.'' 
A more realistic model will have its performance bounded by these two cases.

It can be seen from Figure~\ref{fig_tau} that the optimal $f^*$ is not very sensitive (in this case) to whether or not there is interference between the pico cells. However, Figure~\ref{fig_feasiblef}, indicates that the capacity is more sensitive to the interference. A lower bound of $5.2$ arrivals/sec is provided by the interference case. 

Since it may be difficult to adapt the parameter $f^*$ (the ABS slot size) on a fast time-scale, it is of interest to measure the sensitivity of the range of feasible $f$, as a function of the arrival rate. Figure \ref{fig_feasiblef} depicts the range of ABS values which, when fixed, would allow stable operation,
against arrival rate $\lambda_S$. As can be seen, 
there is only significant sensitivity when the arrival rate is close to capacity.

Under the processor sharing model we also obtained the following results for mean time to send
( transmission time + queueing delays), under no interference. These results depend on the file size only through its
mean. The results for the optimal case are shown in the midlle graph of figure \ref{fig_meandelay} for various overall arrival rates and were obtained using Little's law. The graph to the left shows the same results if the ABS value is taken to be $f = 0.028 < \fstar = 0.106$ and the right graph with $f = 0.203 > \fstar$.
The graphs show that the mean time to send is sensitive to the choice of $f$; compare for example the arrival rates
to attain a mean time to send of 4 seconds in the 3 graphs.

Finally, Figure~\ref{fig_meandelay} illustrates a case in which pico cell $3$ is not saturated at the optimal solution (there is strict inequality in \eqref{eqn_cts_overall_dataconstraint} for pico cell $3$). It follows that capacity can be enhanced if we allow the pico cells to be switched on and off, as in Section~\ref{sec_genschedule}.

 \begin{figure}[t]
    \centering
    \begin{floatrow}
      \ffigbox[\FBwidth]{\caption{Feasible ABS periods against arrival rate $\lambda_S$}\label{fig_feasiblef}}{%
        \includegraphics[width=3in]{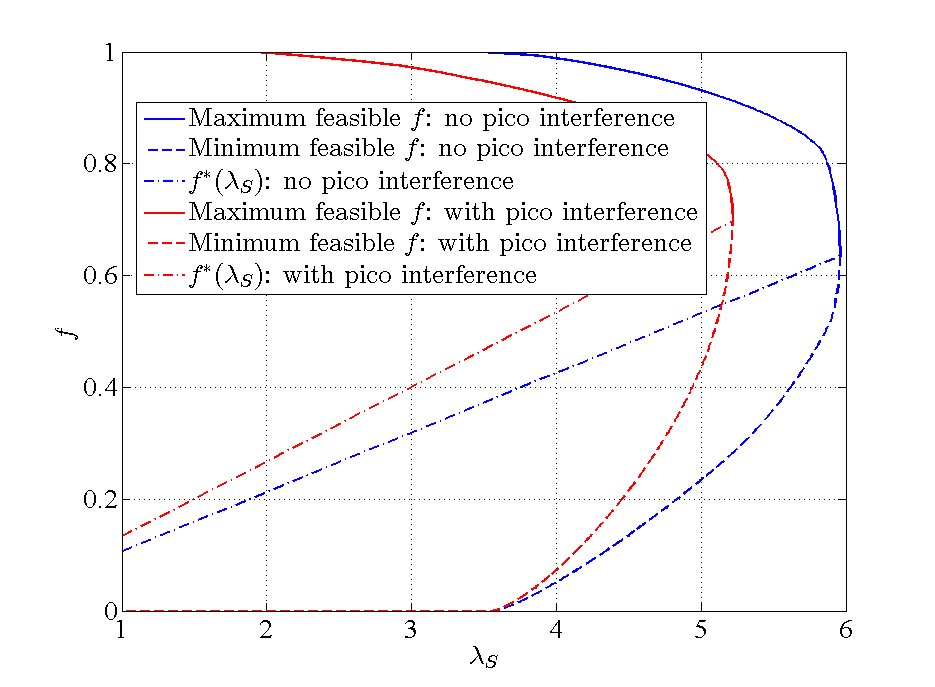}
      }
      \ffigbox[\FBwidth]{\caption{Fraction ABS Time $f$ versus Rate Ratio $\rho$ $L=3$ Pico Cells}\label{fig_meandelay}}{%
        \includegraphics[width=3in]{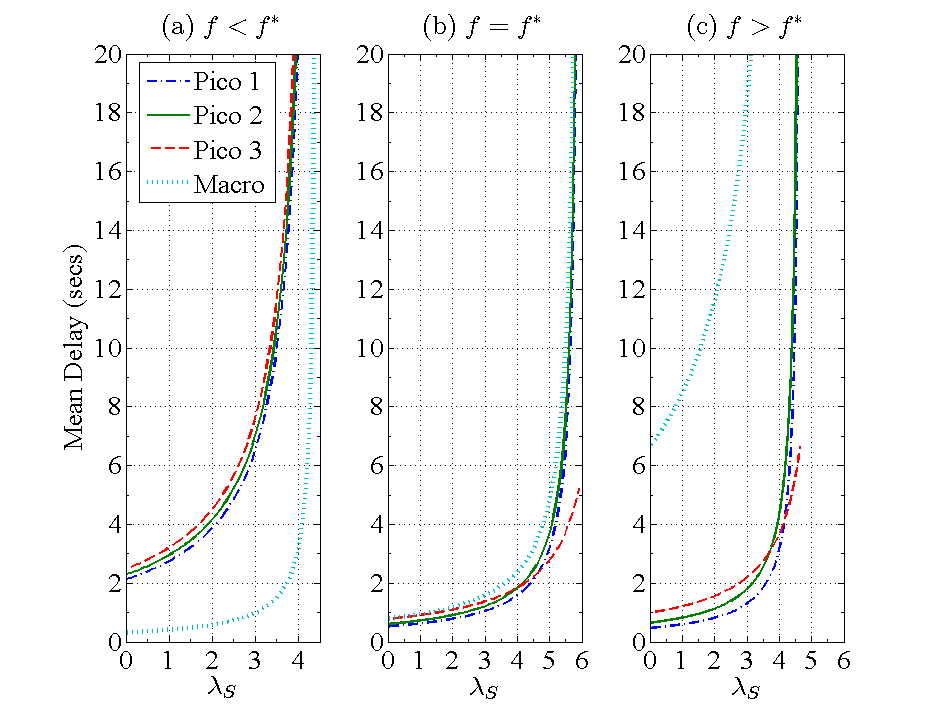}
      }
    \end{floatrow}
\end{figure}

\section{Conclusions}
\label{sec_conc}
This paper has presented a definition of capacity for the down link of a HetNet in terms of the maximum supportable traffic. It can be obtained once the traffic density (probability of arrival at a given location)
as well as the  macro and pico physical rates are given as well as other traffic parameters. The
capacity limit applies irrespective of how the HetNet is scheduled and is subject only to the constraint
that mobiles are cleared from the network.  It is a natural generalization of a criteria to
clear a static network in minimum time.

The capacity can be evaluated as the numerical
solution to a continuous linear program which can be solved efficiently as has been described. Thus
the results of this paper can be used for capacity planning of future networks. Representative
results for transfer delays, numbers of active mobiles and so on, can also be obtained as has
been shown. These results are optimistic in that they suppose the offered traffic given.
They do not take into account the potentially significant gains in transfer delay which an
adaptive scheduler might achieve however. This is a topic for further investigation.

Finally, the characterization of capacity extends to networks with controls such as on-off scheduling.
As indicated an analogous continuous LP applies in such cases.

\appendices

\section{Proof of Theorem \ref{thm_ctsLPsoln}}
\label{appendix_main_proof}
Given any $f \in [0, \fbar]$ we will obtain $\rho_\ell(f), \ell=1,\cdots,L$ so that the Lagrangian is minimized
for the $\xstar_\ell(\xi)$ as stated in the theorem. For each $\ell$ form the convex dual function,
{\small
\begin{equation}
g_\ell(\rho_\ell) = \rho_\ell f + D \int_{A_{\rho_\ell}} \lb \frac{1}{S_\ell(\xi)} - \frac{\rho_\ell}{R_\ell(\xi)}
\rb \xm{d\xi}
\end{equation}
}
which is non-negative and convex. On differentiating under the integral sign, we find that
{\small
\begin{equation}
g'(\rho_\ell) = f - \int_{A_{\rho_\ell}} \frac{D}{R_\ell(\xi)} \xm{d\xi}
\end{equation}
}
which is continuous. Define $\rho_\ell(f)$ to be a minimizer if  $f < \fbar$ in which case
{\small
$$
 f = \int_{A_{\rho_\ell}} \frac{D}{R_\ell(\xi)} \xm{d\xi}
$$
}
otherwise choose $\rho_\ell(f) = 0$. In either case the corresponding $\xstar_\ell$ is feasible.

Now consider maximizing the Lagrangian with the above choice of $\rho_\ell$ as multipliers,
{\small
\begin{eqnarray}
{\cal L} &  = & \int_{{\cal C}_\ell} \frac{x_\ell(\xi)}{S_\ell(\xi)} \xm{d\xi} + \rho_\ell
\lb f - \int_{{\cal C}_\ell} \frac{x_\ell(\xi)}{R_\ell(\xi)} \xm{d\xi} \rb \\
 & = & \rell f + \int_{{\cal C}_\ell} x_\ell(\xi) \lb \frac{1}{S_\ell(\xi)}-\frac{\rell}{R_\ell(\xi)} \rb \xm{d\xi}
 \nonumber
\end{eqnarray}
}
The optimal choice of $x_\ell(\xi) := D$ whenever the expression in brackets is positive, which corresponds
to the choice stated in the theorem and feasible as we have already observed. It follows by the Lagrange Sufficiency Theorem, \cite{Whittle71} that $\xstar_\ell$ is the minimizer for arbitrary $f$. Since this includes the optimal
$\fstar$ the theorem is proved. 
\section{Proof of Lemma~\ref{lem_LLN}}
\label{app_lem_LLN}

For $\xi \in \calC_0$ and $b \in (0,B)$, define $x_{0}^{(a)}$, $y_0^{(a)}$ by
\begin{eqnarray}
x_0^{(a)}(\xi, b) & = & 0 \\
y_{0}^{(a)}(\xi,b) & = & b
\end{eqnarray}
For $\ell=1, 2, \ldots L$, $\xi \in \calC_\ell$, and $b \in (0,B)$, define
\begin{eqnarray}
x_\ell^{(a)}(\xi, b) & = & b I[\rho_\ell(\xi) > \rho_\ell^{(a)}] \\
y_\ell^{(a)}(\xi, b) & = & b I[\rho_\ell(\xi) < \rho_\ell^{(a)}].
\end{eqnarray}

Let $M_T(d\xi,db)$ be the empirical point measure on $S \times (0,\Dbar)$ defined by the points $(\xi_n, D_n)$, $n = 1, 2, \ldots N_T$. Then
\begin{eqnarray}
\frac{1}{T} f_{\ell,T}^{(a)} & = & \int_{\calC_\ell} \int_0^{\Dbar} \frac{x_\ell^{(a)}(\xi,b)}{R_\ell(\xi)} M_T(d\xi,db) \\
\frac{1}{T} Y_{T}^{(a)} & = & \int_{\calC_0} \int_0^{\Dbar} \frac{y_0^{(a)}(\xi,b)}{S_0(\xi)} M_T(d\xi,db) +  \suml \int_{\calC_\ell} \int_0^{\Dbar} \frac{y_\ell^{(a)}(\xi,b)}{S_\ell(\xi)} M_T(d\xi,db)
\end{eqnarray}
By independence of $\xi_n$ and $D_n$, we have
\begin{equation}
M_T(\omega,d\xi,db) \Rightarrow F_B(db) \lambda(d\xi)
\label{eqn_Varaharajan}
\end{equation}
see \cite{Billingsley68}.

Since $x_\ell^{(a)}(\xi, b)$ and $y_\ell^{(a)}(\xi, b)$ are non-negative, bounded and measurable functions on $S \times (0,\Dbar)$, the strong law of large numbers implies
\begin{eqnarray}
\frac{1}{T}f_{\ell,T}^{(a)} & \rightarrow & \int_{\calC_\ell} \int_0^{\Dbar} \frac{x_\ell^{(a)}(\xi,b)}{R_\ell(\xi)} F_B(db) \lambda(d\xi) \\
& = & \int_{\calC_\ell} \frac{D}{R_\ell(\xi)} I[\rho_\ell(\xi) > \rho_\ell^{(a)}] \lambda(d\xi) ~=~f_\ell^{(a)}.
\end{eqnarray}
almost surely. But $\frac{1}{T} f_T^{(a)} \rightarrow \max \frac{1}{T} f_{\ell,T}^{(a)}$ almost surely by continuity of $\max$.

Similar arguments show that
\begin{equation}
\frac{1}{T} Y_T^{(a)} \rightarrow \int_{\calC_0} \frac{D}{S_0(\xi)} \lambda(d\xi) +  \suml \int_{\calC_\ell} \frac{D}{S_\ell(\xi)} I[\rho_\ell(\xi) < \rho_\ell^{(a)}]~a.s.
\end{equation}

\section{Proof of Lemma \ref{lem_LDR}}
\label{app_lemLDR}
Define $m_\ell(n) \doteq \expect{N_{H,T}^{n,\ell}}$
to be the expected number of pico $\ell$ mobiles arriving per unit time with rate ratios falling
in $H_n(\ell)$.
Given $\epsilon > 0$ choose a set of intervals $H_\ell(n), n=1,\cdots,N_P,\ell=1,\cdots,L$ with $N_P$ sufficiently
large so that $m_\ell(n) < \epsilon/2$ for each interval $n$ and pico $\ell$.

Now for  any fixed $0 < \delta < 1/2$ there exists $I_{n,\ell} > 0$
\begin{equation}
\prob{\frac{1}{T} \NTHLn \not \in [(1-\delta) m_\ell(n),(1+\delta) m_\ell(n)]}
  \leq e^{-T I_{n,\ell}}
\label{eqn_Jbnd}
\end{equation}
This  follows from standard large deviation arguments applied
to Poisson variates. Given $\delta,T$ define $K_{\delta,T}$ to be the union of the above events.

From the union bound,
$$
\prob{K_{\delta,T}} \leq \sum_{\ell=1}^L \sum_{n=1}^{N_P} e^{-T I^\delta_{n,\ell}}
\leq N_P L e^{-I^\delta T}
$$
for some $I^\delta > 0$.  It follows from the first Borel-Cantelli lemma
that for any given set of intervals the event $K_{\delta,T}$ will occur
finitely many times with probability 1, there being a last slot
$T_E(\omega) < \infty$ almost surely. We obtain the result on choosing a fixed $0 < \delta < 1$. \qed

\section{Proof of Theorem \ref{thm_ctsLPsoln_modes}}
\label{app_ctsLPsoln_modes}
Let us consider the subproblem in which the $f^\sigma$s are given and each pico $\ell$ minimizes its macro time independently of the others. We drop the pico cell index and restrict all discussion to on-states for which the given cell is active. Consider the dual optimization problem,
\begin{eqnarray}
\min & & \sum_{\sigma} \es \fsgma + D\int Z(\xi)\xm{d\xi} - \int \frac{D}{S(\xi)} \xm{d\xi}
\nonumber \\ 
\mbox{sub} & & Z(\xi) + \frac{\es}{R^\sigma(\xi)} \geq \frac{1}{S(\xi)},~\forall \sigma
\label{eqn_ctsdual} 
\end{eqnarray}
and for which any solution must satisfy $\es \in [0,\overline{\rho}]$.
Clearly it is optimal to take $Z(\xi) = \max_\sigma \ls 1/S(\xi) - \es/R^\sigma(\xi) \rs_+$. If $\fsgma = 0$
it can be deleted from the problem by taking $\es = \overline{\rho}$.

Observe that  (\ref{eqn_ctsdual}) is convex and continuously differentiable in ${\boldsymbol \mu}$. If the optimum is at an interior point, then differentiating we obtain that
$$
\fsgma = \int_{\Aseta} \frac{D}{R^\sigma(\xi)} \xm{d\xi},~\forall \sigma
$$
Suppose that there is a $\sigma$ for which $\mu^\sigma = 0$. Then it follows that $Z(\xi) = 1/S(\xi)$ and since
this is the maximum value which can be taken it follows that $\es = 0, \forall \sigma$. By taking derivatives on
the right,
$$
\fsgma \geq \int \frac{D}{R^\sigma(\xi)} \xm{d\xi}
$$
which shows that no macro time is needed. Additional arguments show that the optimum cannot occur at $\mu^\sigma = \rhoellbar$
since this implies $\fsgma = 0$, which contradicts our assumption that $\fsgma > 0$.

In the primal problem we may take the above solution $\es$ as multipliers and form the Lagrangian,
\begin{equation}
{\mathcal L} =  \int \frac{\lb \sum_\sigma x^\sigma(\xi) - D \rb}{S(\xi)} \xm{d\xi}
+ \sum_\sigma \mu^\sigma \lb \fsgma - \int \frac{x^\sigma(\xi)}{R^\sigma(\xi)} \xm{d\xi} \rb
\label{eqn_primal}
\end{equation}
and set $x^\sigma(\xi) = D$ only on $\Aseta$ and $0$ otherwise. Thus $D$ multiplies the largest
coefficient $\ls \frac{1}{S(\xi)} - \frac{\es}{R^\sigma(\xi)} \rs_+$ over $\sigma$ and the
Lagrangian (\ref{eqn_primal}) is maximal. Since the proposed $x^\sigma$s are feasible they are optimal
by the Lagrange sufficiency Theorem \cite{Whittle71}.

Now consider the overall optimization problem where there are $L$ pico cells so that the values in the vector ${\bf f}$ are
the same for all pico cells. Since the objective is convex and therefore continuous, the optimal choice $\overline{{\bf f}}$ must lie in a compact set and we suppose that the optimal $\overline{{\bf f}}$ is given. We take as Lagrange multipliers the corresponding
$\esell$ which we have shown to exist. For $\sigma$ such that ${\overline f}^\sigma = 0$,
$\esell$ can be taken to be $\rhoellbar$.

Now denote the macro time for any given ${\bf f}$ by $U({\bf f}) := \sum_{\ell=1}^L \Uell({\bf f}).$ If $\Uell >0$ then by standard Lagrangian theory,
$$
- \frac{\partial \Uell}{\partial f^\sigma} = \esell
$$
Hence on differentiating $U(\cdot)$ at $\overline{{\bf f}}$, it follows that
$
\sum_{\ell: \ell \in \sigma} \esell \leq 1$ with equality if $f^\sigma > 0$.

For the final statement, suppose that the constraint \eqref{eqn_ctscon} holds with strict inequality for some $\ell \in \sigma$, in the optimal solution, and let $\barsigma = \sigma \backslash \{\ell\}$. There must be at least one other $j \in \barsigma$ for which \eqref{eqn_ctscon} is tight, or else $f^\sigma$ can be reduced (contradiction). But moving time from $f^\sigma$ to $f^{\barsigma}$ improves the data rates of users in picos in $\barsigma$, because interference for them is less in $\barsigma$ than in $\sigma$. A small amount of exchanged time will not violate \eqref{eqn_ctscon} for $\ell$. But since rates have increased, $f^{\barsigma}$ can be reduced further, contradicting the optimality of $f$.

This completes the proof of the theorem. \qed
\bibliographystyle{IEEETran}

\end{document}